\documentclass[reprint,twocolumn,superscriptaddress,showpacs,floatfix]{revtex4-1}
\usepackage{amsmath}
\usepackage{mathrsfs}
\usepackage{txfonts}
\usepackage{amssymb}
\usepackage{graphicx}
\usepackage{hyperref}
\usepackage{ulem}
\usepackage{color}
 \usepackage{overpic}
 \usepackage{psfrag}
\usepackage{tabularx}
\usepackage{array}
\usepackage{placeins}
\usepackage[marginal]{footmisc}

\newcommand{\PreserveBackslash}[1]{\let\temp=\\#1\let\\=\temp}

\makeatletter

\begin{document}
	
\title{Extracting the number of type-B Goldstone modes and the dynamical critical exponent for a type of scale-invariant states}

\author{Huan-Qiang Zhou}
\affiliation{Centre for Modern Physics, Chongqing University, Chongqing 400044, The People's Republic of
China}

\author{Yan-Wei Dai}
\affiliation{Centre for Modern Physics, Chongqing University, Chongqing 400044, The People's Republic of
China}

\author{Qian-Qian Shi}
\affiliation{Centre for Modern Physics, Chongqing University, Chongqing 400044, The People's Republic of
China}

\author{Ian P. McCulloch}
\affiliation{ Department of Physics, National Tsing Hua University, Hsinchu 30013, Taiwan}
\affiliation{School of Mathematics and Physics, The University of Queensland, St. Lucia, QLD 4072, Australia}
\affiliation{Centre for Modern Physics, Chongqing University, Chongqing 400044, The People's Republic of China}

\author{Murray T. Batchelor}
\affiliation{Mathematical Sciences Institute, The Australian National University, Canberra ACT 2601, Australia}
\affiliation{Centre for Modern Physics, Chongqing University, Chongqing 400044, The People's Republic of China}
	
\begin{abstract}

A generic scheme is proposed to perform a finite-entanglement scaling analysis for scale-invariant states which appear as highly degenerate ground states arising from spontaneous symmetry breaking with type-B Goldstone modes.
This allows us to extract the number of type-B Goldstone modes and the dynamical critical exponent, in combination with a finite block-size scaling analysis, from numerical simulations of quantum many-body systems in the context of tensor network representations.
The number of type-B Goldstone modes is identical to the fractal dimension, thus reflecting an abstract fractal underlying the ground state subspace.
As illustrative examples, we investigate the spin-$s$ Heisenberg ferromagnetic model, the $\rm{SU}(3)$ ferromagnetic model and the $\rm{SO}(4)$ spin-orbital model.
		
\end{abstract}

\maketitle
	
\section{Introduction}
	
In the last decades, much attention has been paid to investigations into quantum critical phenomena~\cite{QPT}.
In particular, significant effort has been made in achieving a long-term goal towards a complete classification of quantum phase transitions and quantum states of matter in one-dimensional quantum many-body systems~\cite{wen,pollmann1}.
Historically, this may be dated back to the early work by Polyakov~\cite{Popkov}, who speculated that scale invariance implies conformal invariance.
This speculation eventually led to the creation of conformal field theory~\cite{Belavin}, thus making it possible to classify all possible critical points in terms of central charge and conformal dimensions.
Given a few counter-examples are known~\cite{Hortacsu, Strominger, LeClair, Zamolodchikov},
it appears to be necessary to launch a systematic search for scale-invariant, but not conformally invariant, quantum states of matter.

Indeed, the presence of scale-invariant, but not conformally invariant, quantum states of matter strongly suggests that the current classification of quantum phase transitions and quantum states of matter is far from complete, even for those relevant to spontaneous symmetry breaking (SSB)~\cite{anderson}.
As it turns out, highly degenerate ground states arising from SSB with type-B Goldstone modes (GMs) are scale-invariant~\cite{FMGM,golden,LLspin1}, with the ${\rm SU}(2)$ Heisenberg ferromagnetic model being a paradigmatic example.
In fact the highly degenerate ground states admit an exact singular value decomposition, thus exhibiting self-similarities underlying the ground state subspace.
In other words, an abstract fractal is revealed, living in a Hilbert space, which may be characterized in terms of the fractal dimension, first introduced by Castro-Alvaredo and Doyon~\cite{doyon} for  the ${\rm SU}(2)$ Heisenberg ferromagnetic states.
A remarkable fact is that the fractal dimension may be identified with the number of type-B GMs~\cite{FMGM,golden}, thus unveiling a deep connection between scale-invariant states and the counting rule of the GMs~\cite{Watanabe}.
The establishment of  the counting rule is largely based on an insightful observation made by Nambu~\cite{nambu}, culminating in the classification of type-A and type-B GMs.
	
In addition, our current understanding of quantum critical phenomena has been reshaped from the novel perspective of quantum information science~\cite{nielsen,amico}.
In particular, the entanglement entropy is demonstrated to be a powerful means for characterizing quantum phase transitions~\cite{vidal,Korepin,cardy,zhou}.
This in turn motivated the development of powerful tensor network representations to simulate quantum many-body systems~\cite{itebd,idmrg,Orus,ipeps,frank,corboz,czarnik}.
As a by-product, a finite-entanglement scaling analysis is developed to replace a finite-size scaling analysis, as advocated~\cite{tagliacozzo,wanghl,pollmann} for conformally invariant quantum states, which allows to extract central charge from numerical simulations in the infinite Matrix Product State (iMPS) representation  for one-dimensional quantum many-body systems.
A natural question arises as to whether or not there is a parallel between scale-invariant quantum states and conformally invariant quantum states.
That is, a generic scheme to perform a finite-entanglement scaling analysis is needed for scale-invariant quantum states.
	
This work aims to address this question for scale-invariant states, which appear to be highly degenerate ground states arising from SSB with type-B GMs.
This allows us to extract the number of type-B GMs and the dynamical critical exponent, in combination with a finite block-size scaling analysis, from numerical simulations of quantum many-body systems in the context of tensor network representations.
As recently argued~\cite{FMGM,golden}, the number of type-B GMs is identical to the fractal dimension, thus reflecting an abstract fractal underlying the ground state subspace.
As illustrative examples, we investigate the spin-$s$ Heisenberg ferromagnetic model, the $\rm{SU}(3)$ ferromagnetic model and the $\rm{SO}(4)$ spin-orbital model.

\section{Finite-entanglement scaling for scale-invariant states}

Let us consider a quantum many-body system described by Hamiltonian $\mathscr{H}$ on a lattice.
If the Hamiltonian $\mathscr{H}$ possesses the symmetry group $G$, which is spontaneously broken into $H$,  then the counting rule  is established as $N_A+2N_B=N_{BG}$~\cite{Watanabe}, where $N_A$ and $N_B$ are the numbers of type-A and type-B GMs, and $N_{BG}$ is equal to the dimension of the coset space $G/H$.
Here and hereafter, we focus on a one-dimensional quantum many-body system, with $L$ being the system size.
According to the Mermin-Wagner-Coleman theorem~\cite{mwc}, no type-A GM survives in one spatial dimension.
Hence the number $N_A$ of type-A GMs must be zero.

Suppose the system is partitioned into a block $B$ and its environment $E$, with the block consisting of $n$ (contiguous) lattice sites, and the other $L-n$ lattice sites constituting the environment $E$.
As demonstrated in Refs.~\cite{FMGM,golden},  for any non-zero fillings $f_\alpha$\; ($\alpha =1, \cdots, R$, with $R$ beng the rank of $G$, if it is semisimple), the block entanglement entropy $S_f(n)$ scales logarithmically with the block size $n$	 in the thermodynamic limit $L \rightarrow \infty$:
	\begin{equation}
	S_f(n)= \frac {N_B}{2} \ln n +S_{f0},
	\label{Scale}
	\end{equation}
where $S_{f0}$ is an additive non-universal constant.
Combining with a field-theoretic prediction made by Castro-Alvaredo and  Doyon~\cite{doyon},  one is able to identify  the number of type-B GMs
with the fractal dimension $d_f = N_B$~\cite{FMGM,golden}.
Such a logarithmic scaling behavior of the block entanglement entropy $S_f(n)$ provides an efficient way to extract the number of type-B GMs, or equivalently, the fractal dimension from numerical simulations of quantum many-body systems in the context of the iMPS representation~\cite{idmrg}.
However, this requires us to ensure that the simulation results be accurate, with the bond dimension $\chi$ being extremely large.
We have described a subroutine to efficiently evaluate the block entanglement entropy $S_f(n)$ in Section A of the Supplementary Material (SM).
In this way we are able to perform a finite block-size scaling analysis of the entanglement entropy $S_f(n)$ for scale-invariant states, which appear to be highly degenerate ground states for quantum many-body systems undergoing SSB with type-B GMs (for more details, cf. Section B of the SM).

A more convenient way is to develop a finite-entanglement scaling analysis for scale-invariant states in the context of the iMPS representation, in parallel to conformally invariant states~\cite{tagliacozzo,wanghl,pollmann}.
For this purpose, we turn to the entanglement entropy $S_f(\chi)$ for the semi-infinite chain instead of a finite-size block, which is defined as
$S_f(\chi) = -\sum _\alpha \Lambda^2_\alpha \ln \Lambda^2_\alpha$,
in terms of the singular values $\Lambda_\alpha$ ($\alpha = 1, \ldots, \chi$, with $\chi$ being the bond dimension).
To achieve this goal, we remark that, for a finite-size block, the time $\tau$ taken for a local disturbance to propagate through the entire block in a coherent way scales linearly with $n$:  $ \tau \sim n$.
Taking into account the dispersion relation $\omega\sim k^{z}$,
with $z$ being the dynamical critical exponent, we have $\xi\sim \tau^{z}$, where $\xi$ is the correlation length.
Hence it is plausible to replace $n$ by $\xi^{1/z}$.
Accordingly,  for any nonzero fillings $f_\alpha (\alpha =1, \cdots, R)$, the entanglement entropy $S_f(\chi)$ for the semi-infinite chain takes the form
	\begin{equation}
		{S_f(\chi)=\frac{N_B}{2z} \ln \xi(\chi)+S_{f0}}(\chi).
		\label{S}
	\end{equation}
Here $S_{f0}(\chi)$ is an additive non-universal constant.
We remark that the correlation length $\xi(\chi)$ scales as  $\xi\sim\chi^{\kappa}$, with $\kappa$ being the finite-entanglement scaling exponent, introduced in Ref.~\cite{tagliacozzo} for conformally invariant states.
	
With the above discussions in mind, we are able to extract the number of type-B GMs, or equivalently, the fractal dimension,
from the iMPS representation for scale-invariant states, if the dynamical critical exponent is known, and vice versa.
Here we remark that an alternative way to extract the dynamical critical exponent $z$ from simulation results of the infinite Density Matrix Renormalization Group (iDMRG) algorithm~\cite{idmrg} is described in Section C of the SM.

\section{Illustrative examples}
	
To illustrate the generic scheme we focus on three fundamental models, which are chosen as typical examples to exhibit SSB with type-B GMs.

The first model is the spin-$s$ Heisenberg ferromagnetic model described by the Hamiltonian
	\begin{equation}
		\mathscr{H}=-\sum_{j}\mathbf{S}_{j} \cdot \mathbf{S}_{j+1},
		\label{ham1}
	\end{equation}
where $\mathbf{S}_j=(S_{j}^{x},S_{j}^{y},S_{j}^{z})$ denotes the spin-$s$ operator at site $j$.
The model possesses the symmetry group $\rm {SU} (2)$, with the generators being $S^x = \sum_j S_j^x, S^y = \sum_j S_j^y$ and $S^z = \sum_j S_j^z$.
In this case, SSB occurs from $\rm {SU} (2)$ to $\rm {U} (1)$, with the number of type-B GMs $N_B=1$~\cite{Watanabe}.
	
The second model is the $\rm{SU}(3)$ spin-1 ferromagnetic model described by the Hamiltonian
	\begin{equation}
	\mathscr{H} =- \sum_{j}
		\left[ (\mathbf{S}_{j} \cdot \mathbf{S}_{j+1})
		+ (\mathbf{S}_{j} \cdot \mathbf{S}_{j+1})^2
		\right],
		\label{ham2}
	\end{equation}
where $\mathbf{S}_j=(S_{j}^{x},S_{j}^{y},S_{j}^{z})$ are spin-$1$ operators at site $j$.
Note that this model appears as a special case of the well-studied spin-1 bilinear-biquadratic model, with its peculiarity being that the Hamiltonian (\ref{ham2}) is exactly solvable by means of the Bethe ansatz~\cite{Sutherland}.
The model possesses the symmetry group $\rm{SU(3)}$, with the generators being realized in terms of the spin-1 operators
$K_{\alpha}=\sum_{j}K_{\alpha}^j$
$(\alpha= 1, 2, \ldots, 8)$, where
$K_1=1/2\sum_{j}S_j^x$, $K_2=1/2\sum_{j}S_j^{y}$, $K_3=1/2\sum_{j}S_j^z$,
$K_4=1-3/2\sum_{j}(S_j^z)^2$, $K_5=1/2\sum_{j}({(S_j^x)}^2-{(S_j^y)}^2)$,
$K_6=1/2\sum_{j}(S_j^yS_j^z+S_j^zS_j^y)$, $K_7=1/2\sum_{j}(S_j^zS_j^x+S_j^xS_j^z)$ and
$K_8=1/2\sum_{j}(S_j^xS_j^y+S_j^yS_j^x)$.
In this case, the number of type-B GMs $N_B=2$,
given SSB occurs from $\rm{SU}(3)$ to $\rm{U}(1) \times \rm{U}(1)$~\cite{FMGM}.

The third model is the ${\rm SO}(4)$ spin-orbital model described by the Hamiltonian~\cite{So4}
	\begin{equation}
			\mathscr{H}=-\sum_{j}(\zeta+\mathbf{S}_j \cdot \mathbf{S}_{j+1})(\zeta+\mathbf{T}_j \cdot \mathbf{T}_{j+1}),\label{ham3}
	\end{equation}
where  $\mathbf{S}_j=(S_{j}^{x},S_{j}^{y},S_{j}^{z})$ are spin-$1/2$ operators and $\textbf{T}_j=(T_{j}^x,T_{j}^y,T_{j}^z)$ are orbital pseudo-spin 1/2 operators at site $j$.
Here we restrict ourselves to the two particular points $\zeta=3/4$ and $\zeta=\infty$, both of which share the same ground state subspace.
Actually, the model (\ref{ham3}), with  $\zeta=3/4$ and $\zeta=\infty$, may be recognized as the model studied by Kolezhuk and Mikeska~\cite{kolezhuk}, with the opposite sign, and
a model consisting of two decoupled ${\rm SU(2)}$ spin-$1/2$ Heisenberg ferromagnetic chains.
The model possesses  the  symmetry group ${\rm SO(4)}$, isomorphic to ${\rm SU(2)} \times {\rm SU(2)}$, with the generators of the two copies of  ${\rm SU(2)}$ being $S^x=\sum_jS_{j}^x$, $S^y=\sum_jS_{j}^y$,  and $S^z=\sum_jS_{j}^z$, and $T^x=\sum_jT_{j}^x$, $T^y=\sum_jT_{j}^y$ and $T^z=\sum_jT_{j}^z$.
In this case, SSB occurs from ${\rm SO(4)} $ to ${\rm U(1)}\times {\rm U(1)}$, with the number of type-B GMs $N_B=2$.

\section{Simulation results}

In order to simulate the three selected models (\ref{ham1}), (\ref{ham2}) and (\ref{ham3}), one needs to develop the iDMRG algorithm~\cite{idmrg}, with ${\rm U}(1)$, $\rm{U}(1) \times \rm {U}(1)$ and $\rm{U}(1) \times \rm {U}(1)$ being implemented as a symmetry group, respectively.
This is necessary, since we have to target a specific ground state with a given filling.
In this sense, the iDMRG algorithm is the method of choice, which is able to efficiently produce the iMPS representation for one of the highly degenerate ground states arising from SSB with type-B GMs.

	\begin{figure}
		\includegraphics[width=0.32\textwidth]{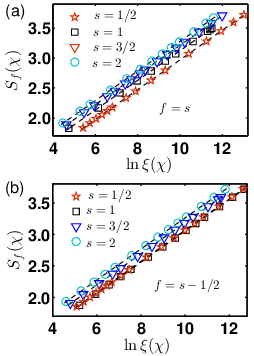}
		\caption{The entanglement entropy $S_f(\chi)$ versus $\ln \xi(\chi)$ for the spin-$s$ Heisenberg ferromagnetic model
       (\ref{ham1}) with $s$ being $1/2, 1, 3/2$ and 2. The different fillings (a) $f=s$ and (b) $f=s-1/2$ have been chosen.
       The bond dimension $\chi$ ranges from $20$ to $160$.  }
		\label{SandCsu2}
	\end{figure}

	\begin{figure}
		\includegraphics[width=0.3\textwidth]{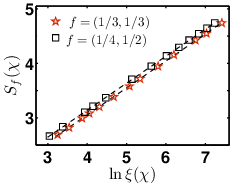}
		\caption{The entanglement entropy $S_f(\chi)$ versus $\ln \xi (\chi)$ for the $\rm{SU}(3)$ ferromagnetic model
       (\ref{ham2}). Here we have chosen the fillings $f=(1/3, 1/3)$ and $f=(1/4, 1/2)$, with the bond dimension $\chi$ ranging from $80$ to $1000$.}
		\label{SandCsu3}
	\end{figure}

	\begin{figure}
		\includegraphics[width=0.3\textwidth]{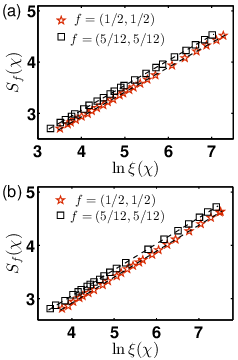}
		\caption{The entanglement entropy  $S_f(\chi)$ versus $\ln \xi (\chi)$ for the ${\rm SO}(4)$ spin-orbital model
        (\ref{ham3}) with  $\zeta=3/4$ and $\zeta=\infty$.
		Here we have chosen the fillings $f=(1/2, 1/2)$ and $f=(5/12,5/12)$, with
			the bond dimension $\chi$ ranging from $100$ to $1000$.}
		\label{SSandCso4}
	\end{figure}

In Fig.~\ref{SandCsu2}, we plot the entanglement entropy $S_f(\chi)$ versus $\ln \xi(\chi)$ for the spin-$s$ Heisenberg ferromagnetic model (\ref{ham1})
with $s = 1/2, 1, 3/2$ and $2$.
Here we have chosen different fillings: (a) $f=s$ and (b) $f=s-1/2$, with the bond dimension $\chi$
ranging from $20$ to $160$.
To this end, the unit cell in the iMPS representation consists of  two lattice sites.
For $f=s$, the number of type-B GMs is extracted to be $N_B=0.9924, 0.9904, 0.992$ and $0.9912$ for $s=1/2, 1, 3/2$ and $2$, respectively.
The relative error is less than $1\%$, if we adopt the value of the dynamical critical exponent to be $z=2$, as predicted from the conventional spin wave theory.
Conversely, the other way around is to extract the dynamical critical exponent if we adopt $N_B=1$.
The dynamical critical exponent is extracted to be $z=2.0153, 2.0194, 2.0161$ and $2.0178$ for $s=1/2, 1, 3/2$ and $2$, respectively.
The relative error is less than $1\%$.
Similarly, for $f=s-1/2$, the number of type-B GMs is extracted to be $N_B=0.9928, 0.992, 0.9904$ and $0.9916$ for $s=1/2, 1, 3/2$ and $2$, respectively.
Then a relative error is less than $1\%$, if we adopt the dynamical critical exponent value $z=2$.
Conversely, the dynamical critical exponent is extracted to be $z=2.0145, 2.0161, 2.0194$ and $2.0169$ for $s=1/2, 1, 3/2$ and $2$, respectively, with the relative error again less than $1\%$.

In Fig.~\ref{SandCsu3}, we plot the entanglement entropy $S_f(\chi)$ versus $\ln \xi(\chi)$ for the $\rm{SU}(3)$ ferromagnetic model (\ref{ham2}).
The filling factors are chosen to be  $f=(1/3, 1/3)$ and $f=(1/4, 1/2)$, with the bond dimension $\chi$ ranging from $80$ to $1000$.
Here the unit cell in the iMPS representation consists of  three and four lattice sites, respectively.
The number of type-B GMs is extracted to be $N_B=1.9804$ for $f=(1/3, 1/3)$ and $1.9816$  for $(1/4, 1/2)$, with the relative error being less than $1\%$, if we adopt the dynamical critical exponent $z=2$, as predicted from the conventional spin wave theory.
Conversely, adopting the value $N_B=2$, the dynamical critical exponent is extracted to be $z=2.0198$ for $f=(1/3, 1/3)$ and $2.0186$ for $(1/4, 1/2)$, with again the relative error less than $1\%$.

Fig.~\ref{SSandCso4} shows plots of the entanglement entropy $S_f(\chi)$ versus $\ln \xi(\chi)$ for the ${\rm SO}(4)$ spin-orbital model (\ref{ham3}),
with (a) $\zeta=3/4$ and (b) $\zeta=\infty$.
Here we have chosen the fillings $f=(1/2, 1/2)$ and $f=(5/12,5/12)$, with
the bond dimension $\chi$ ranging from $100$ to $1000$.
The unit cell in the iMPS representation consists of four and six lattice sites, respectively.
If we adopt the dynamical critical exponent $z=2$, as predicted from the conventional spin wave theory,
the best linear fit is exploited to estimate the number of type-B GMs as $N_B=1.9296$ and $1.9468$ in Fig.~\ref{SSandCso4}(a), with relative error less than $4\%$, and  $N_B=1.9512$ and $1.9524$ in Fig.~\ref{SSandCso4}(b), with relative error less than $3\%$.
Here if we adopt $N_B=2$, the dynamical critical exponent is extracted to be
 $z=2.073$ and $2.059$ in Fig.~\ref{SSandCso4}(a) and $z=2.05$ and $2.0488$ in Fig.~\ref{SSandCso4}(b), with in each case a relative error less than $4\%$.

\section{Summary}
	
A generic scheme to perform a finite-entanglement scaling analysis has been put forward for highly degenerate ground states arising from SSB with type-B GMs, which are scale-invariant, but not conformally invariant.
Extensive numerical simulations have been performed for the three selected models -- the spin-$s$ Heisenberg ferromagnetic model, the $\rm{SU}(3)$ ferromagnetic model and the
$\rm{SO}(4)$ spin-orbital model. Actually, the number of type-B GMs $N_B$ may be reliably extracted from finite block-size scaling, as long as the bond dimension $\chi$ is large enough, within a reasonable accuracy.
A detailed exposition for a finite block-size scaling analysis to extract the number of type-B GMs has been described in Section D of the SM.
This in turn allows us to extract the dynamical critical exponent from finite-entanglement scaling.
	
\acknowledgements
	
We appreciate insightful discussions with John Fjaerestad about the counting rule for the GMs.

\end{document}


\title{Supplementary Material for ``Extracting the number of type-B Goldstone modes and the dynamical critical exponent for a type of scale-invariant states"}

\author{Huan-Qiang Zhou}
\affiliation{Centre for Modern Physics, Chongqing University, Chongqing 400044, The People's Republic of
China}

\author{Yan-Wei Dai}
\affiliation{Centre for Modern Physics, Chongqing University, Chongqing 400044, The People's Republic of
China}

\author{Qian-Qian Shi}
\affiliation{Centre for Modern Physics, Chongqing University, Chongqing 400044, The People's Republic of
China}

\author{Ian P. McCulloch}
\affiliation{ Department of Physics, National Tsing Hua University, Hsinchu 30013, Taiwan}
\affiliation{School of Mathematics and Physics, The University of Queensland, St. Lucia, QLD 4072, Australia}
\affiliation{Centre for Modern Physics, Chongqing University, Chongqing 400044, The People's Republic of China}

\author{Murray T. Batchelor}
\affiliation{Mathematical Sciences Institute, The Australian National University, Canberra ACT 2601, Australia}
\affiliation{Centre for Modern Physics, Chongqing University, Chongqing 400044, The People's Republic of China}

\maketitle

\section*{A.~~ A subroutine to efficiently evaluate the block entanglement entropy $S_f(n)$}
	\label{appA}
	In order to extract the number of type-B GMs $N_B$, or equivalently, the fractal dimension $d_f$, from a
	finite-size scaling analysis of the block entanglement entropy $S_f(n)$, a subroutine is developed to  efficiently evaluate
	the block entanglement entropy $S_f(n)$.
	
\begin{figure}[ht]
		\includegraphics[width=0.42\textwidth]{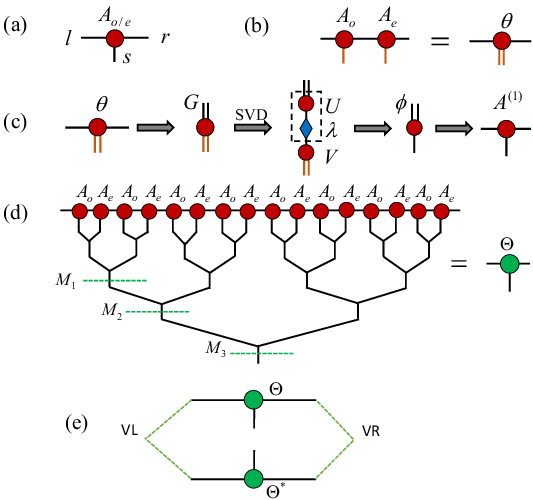}
		\caption{(a) A single tensor  $A_{o/e}$ in the iMPS representation. Here $o/e$ refers to an odd/even site,
			$l$ and $r$ are the bond
			indices: $l,r=1,...,\chi$, with $\chi$ the bond dimension, and $s$ the physical index: $s=1,...,d$, with $d$ the
			dimension of the local Hilbert space.
			(b) Contraction of two tensors $A_o$ and $A_e$ to form one single tensor $\theta$.
			(c) Reshape $\theta$ into a matrix $G$, labeled by the
			combined bond indices and the combined  physical indices, perform a singular value decomposition for $G$, and truncate
			the physical indices to ensure that $|\lambda_M/\lambda_1| < \epsilon$, where $\lambda_1, \cdots, \lambda_M$ are the singular
            values,
			with $M$ the number of states to be kept, and $\epsilon$ a preset precision. Afterwards, discard $V$ and contract $U$ and
			$\lambda$ to form one single tensor $\phi$ and reshape it into one single tensor $A^{(1)}$.
			(d)  Repeat this process until the block size is reached, thus leading to one single tensor $\Theta$.
			(e) An approximation to  the reduced density matrix  for a block may be evaluated from $\Theta$. Here we have chosen the block
			size $n=16$ for illustration.
			Hence, a sequence of successive truncation consists of three different types of contractions, with $M_1$,  $M_2$ and  $M_3$
            denoting the number of
			states to be kept after each successive truncation.		}
		\label{Method}
	\end{figure}

	We assume that  a ground state $|\psi\rangle$ in the iMPS representation is invariant under the two-site translation.  Hence, it
	is represented by two tensors  $A_o$ and $A_e$, as shown in Fig.~\ref{Method}(a), where
	$l$ and $r$ are the bond indices: $l,r=1,...,\chi$, with $\chi$ being the bond dimension, and $s$ is the physical index: $s=1,...,d$,
	with $d$ being the dimension of the local Hilbert space.

\begin{figure}[ht]
		\includegraphics[width=0.32\textwidth]{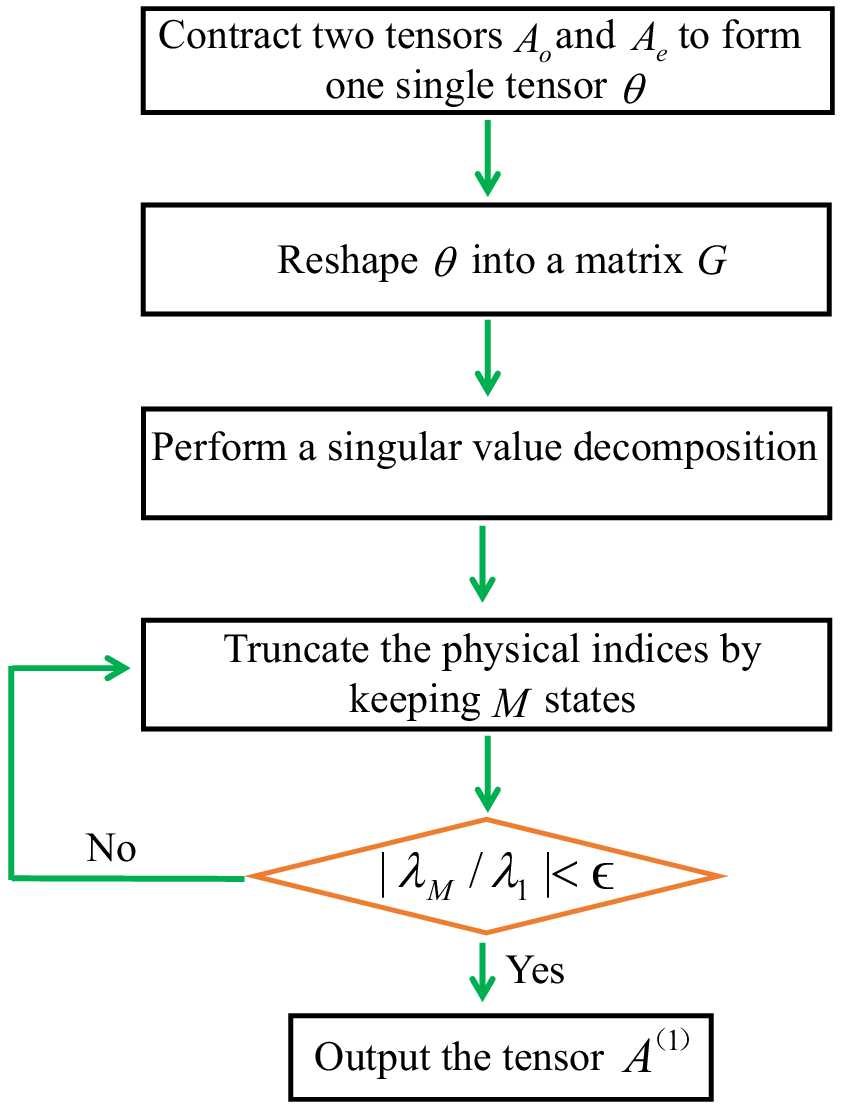}
		\caption{A flow chart to contract two tensors $\Gamma_A$ and $\Gamma_B$ to form one single tensor $\theta$, reshape $\theta$ into a
            matrix  $G$, labeled by the
			combined bond indices and the combined  physical indices,   perform a singular value decomposition for $G$, and truncate
			the physical indices to ensure that $|\lambda_M/\lambda_1| < \epsilon$, where $\lambda_1, \ldots, \lambda_M$ are the singular
           values, with $M$ being the number of states to be kept, and $\epsilon$ is a preset precision.}
		\label{flowchart}
	\end{figure}

	The subroutine to efficiently evaluate the block entanglement entropy $S_f(n)$ consists of three steps, as shown in parts (b), (c) and
    (d) of
	Fig.~\ref{Method}:
	First, contract two tensors $A_o$ and $A_e$ to form one single tensor $\theta$, reshape $\theta$ into a matrix  $G$, labeled by the
	combined bond indices and the combined  physical indices,   perform a singular value decomposition for $G$, and truncate
	the physical indices, with a
	flow chart being shown in Fig.~\ref{flowchart}.
	Second, repeat the above process for a newly-formed tensor $A^{(1)}$ until the block size is reached,
	thus leading to a tensor $\Theta$. Third,  an approximation to the reduced density matrix $\rho_{n}$ may be evaluated from the tensor $\Theta$.
	
	The subroutine is tested for the $\rm {SU} (2)$ spin-$1/2$ ferromagnetic states, with bond dimension
	$\chi=50$.
	For comparison, the block entanglement entropy $S_f(n)$ takes the value $2.065418553$  if no truncation is performed. Hence the subroutine yields a
	reliable result, with
	a relative error being less than $3.6 \times 10^{-7}$.
	See Table~\ref{table0}.

	\begin{table}[ht]
		\renewcommand\arraystretch{2}
				\caption{The block entanglement entropy $S_f(n)$, with block size $n=16$ and bond dimension $\chi=50$,
				is evaluated for the $\rm {SU} (2)$ spin-$1/2$ ferromagnetic states.
			$\epsilon$ denotes a preset precision for performing repeated truncation. }
	        \vskip 2mm
		\begin{tabular}{ccccccccccccc}
			\hline
			\hline
			\begin{minipage}{0.8cm} $\epsilon$ \end{minipage}
			& \begin{minipage}{0.7cm} $M_1$ \end{minipage} &
			\begin{minipage}{0.7cm} $M_2$ \end{minipage} &
			\begin{minipage}{0.7cm} $M_3$ \end{minipage} &
			\begin{minipage}{1.7cm} $S_f(n)$ \end{minipage}&
			\begin{minipage}{0.9cm} relative error\end{minipage} \\
			\hline
			\begin{minipage}{0.8cm} $10^{-3}$ \end{minipage}
			& 16 &70& 160 & 2.065417803& $3.6 \times 10^{-7}$ \\
			\begin{minipage}{0.8cm} $10^{-4}$ \end{minipage}
			& 16 &90& 250 & 2.065418523 & $1.5 \times 10^{-8}$ \\
			\begin{minipage}{0.8cm} $10^{-5}$ \end{minipage}
			& 16 &120& 320& 2.065418552 & $5.7 \times 10^{-10}$ \\
			\hline
			\hline
		\end{tabular}
		\label{table0}
	\end{table}

\section*{B.~~ Finite block-size scaling of the entanglement entropy $S_f(n)$}
	
	The subroutine to efficiently evaluate the block entanglement entropy $S_f(n)$, described in Subsection A, is exploited to perform a finite block-size scaling analysis of the entanglement entropy $S_f(n)$~\cite{FMGM0,golden0} for the three selected models  -- the spin-$s$ Heisenberg ferromagnetic model, the $\rm{SU}(3)$ ferromagnetic model and the $\rm{SO}(4)$ spin-orbital model -- in the context of the iDMRG algorithm~\cite{idmrg0}, with a fixed  value of the bond dimension.
	A more detailed exposition on the extraction of the number of type-B GMs from different values of the bond dimension $\chi$ is outlined in Section D.
	
\begin{figure}
		\includegraphics[width=0.3\textwidth]{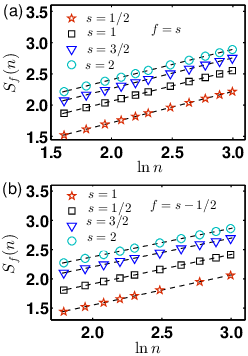}
		\caption{The entanglement entropy $S_f(n)$ versus $\ln n$ for the $\rm {SU} (2)$ spin-$s$ Heisenberg ferromagnetic
			model~(1),
			with $s=1/2, 1, 3/2$ and 2. We have chosen different fillings: (a) $f=s$ and (b) $f=s-1/2$, with
			bond dimension $\chi=160$.}
		\label{SandNSU2}
	\end{figure}

	In Fig.~\ref{SandNSU2}, we plot the block entanglement entropy $S_f(n)$ versus $\ln n$ for the spin-$s$ Heisenberg ferromagnetic model~(3), with $s=1/2, 1, 3/2$ and 2, as a simulation result of the $\rm {U}(1)$ iDMRG algorithm in the iMPS representation for different fillings $f$.
	For $f=s$, the number of type-B GMs is extracted to be $N_B=1.002, 0.982$, $0.9786$ and $0.97$ for $s=1/2, 1, 3/2$ and $2$, respectively, where the block size ranges from $n=5$ to $n=20$. The relative error is less than $3\%$, compared to the exact value $N_B=1$.
	For  $f=s-1/2$, the number of type-B GMs is extracted to be  $N_B=1.0298, 1.0052, 0.9804$ and $0.9722$ for $s=1/2, 1, 3/2$ and $2$, respectively, with the block size ranging from $n=6$ to $n=20$. The relative error is less than $3\%$ compared to the exact value $N_B=1$.
	
\begin{figure}
		\includegraphics[width=0.3\textwidth]{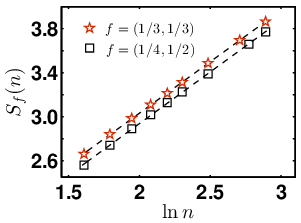}
		\caption{The  entanglement entropy $S_f(n)$ versus $\ln n$ for the $\rm{SU}(3)$
			ferromagnetic model~(4). The different fillings are as indicated, with
			bond dimension $\chi=300$.
		}
		\label{SandNSU3}
	\end{figure}

	In Fig.~\ref{SandNSU3}, we plot the block entanglement entropy $S_f(n)$ for the $\rm{SU}(3)$ ferromagnetic model~(4), as a simulation result of the $\rm {U}(1) \times \rm {U}(1)$ iDMRG algorithm in the iMPS representation for  $f=(1/3, 1/3)$ on the three-site unit cell and for $f=(1/4, 1/2)$ on the four-site unit cell.
	The best linear fit is exploited to estimate the number of type-B GMs $N_B=1.8702$ for $f=(1/3, 1/3)$ and $1.8772$ for $(1/4, 1/2)$,
   with relative error less than $6.5\%$, compared to the exact value $N_B=2$.
	The block size ranges from $n=5$ to $n=18$.

	\begin{figure}
		\includegraphics[width=0.3\textwidth]{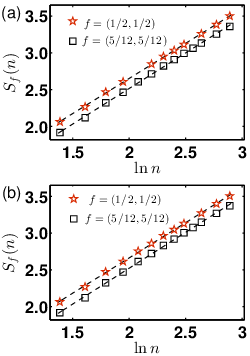}
		\caption{The  entanglement entropy $S_f(n)$ versus $\ln n$
			for the $\rm{SO}(4)$ spin-orbital model~(5) with (a) $\zeta=3/4$ and (b) $\zeta=\infty$.
			The different fillings are as indicated, with	bond dimension $\chi=400$.
		}
		\label{SandNSo4}
	\end{figure}

	In Fig.~\ref{SandNSo4}, we plot the block entanglement entropy $S_f(n)$ versus $\ln n$ for the ${\rm SO}(4)$ spin-orbital
	model (5) with (a) $\zeta=3/4$ and (b) $\zeta=\infty$, as a simulation result of the $\rm{U}(1)\times \rm{U}(1)$ iDMRG algorithm
	in the iMPS representation for  $f=(1/2, 1/2)$
	on the four-site unit cell and for  $f=(5/12, 5/12)$ on the six-site unit cell.
	For $\zeta=3/4$,
	the best linear fit is exploited to estimate the number of type-B GMs $N_B=1.9104$ for  $f=(1/2, 1/2)$ and $1.9352$ for $(5/12,
	5/12)$,
	with relative error less than $4.5\%$, compared to the exact value $N_B=2$.
	For $\zeta=\infty$,
	the best linear fit is exploited to estimate the number of type-B GMs $N_B=1.9212$ for $f=(1/2, 1/2)$ and $1.9544$  for $(5/12,
	5/12)$, with relative error less than $4\%$, compared to the exact value $N_B=2$.
	The block size ranges from $n=4$ to $n=18$.

	\section*{C.~~ The dynamical critical exponent $z$ from two types of correlation lengths}
	
\begin{figure}
		\includegraphics[width=0.44\textwidth]{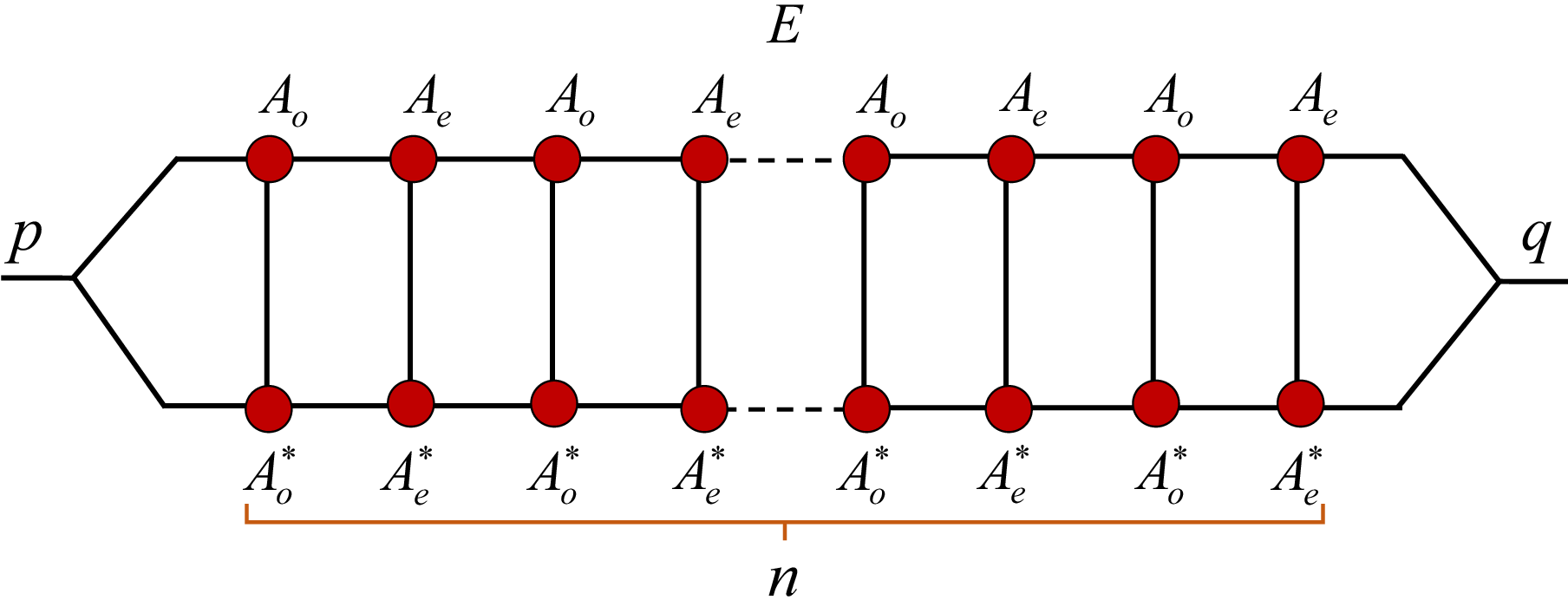}
		\caption{A diagrammatical representation of the transfer matrix $E$, consisting of $n$ contiguous lattice sites, with $n$ even.
			Here $p = (p_1,\ldots,p_R)$ and $q = (q_1,\ldots,q_R)$ label the left and right bonds, with $R$ the rank of a (semisimple)
           symmetry group $G$. The SSB pattern is from a symmetry group $G$ to a residual symmetry group $H$: $G \rightarrow H$.}
		\label{EE}
	\end{figure}

	The correlation length $\xi$ is defined  as $\xi=1/\ln|\Lambda_1/\Lambda_2|$  with $\Lambda_1$ and $\Lambda_2$  the largest and
	second largest eigenvalues of the transfer matrix $E$ for a block consisting of $n$ contiguous lattice sites~\cite{baxter}, as schematically shown in
	Fig.~\ref{EE}.
	We exploit $p$ and $q$ to label the left and right bonds, where $p = (p_1,\ldots,p_R)$ and $q = (q_1,\ldots,q_R)$, with $R$ being the rank of a (semisimple) symmetry group $G$. Here, the SSB pattern is from a symmetry group $G$ to a residual symmetry group $H$: $G \rightarrow H$. Hence the entries of the transfer matrix $E$ vanish unless $p=q$. That is, the  transfer matrix $E$ is in a block-diagonal form.
	
	With this fact in mind, one may introduce two distinct types of correlation lengths, each of which in principle constitutes an infinite sequence. One type is $\xi_{u,q}$, defined as $\xi_{u,q}= 1/\ln (\Lambda_{1,q}/\Lambda_{2,q})$, which is restricted to a given sector labeled by $q$.
Here all sectors are allowed, including the first sector, labelled by $q=(0,\ldots,0)$.
The other type is $\xi_{q}$ ($q \neq (0,\ldots,0)$),  defined as $\xi_{q}= 1/\ln (\Lambda_{1,0}/\Lambda_{1,q})$, which involves two distinct sectors, labelled by $(0,\ldots,0)$ and $q$.
$\Lambda_{1,q}$ and $\Lambda_{2,q}$ denote the largest and second largest eigenvalues of the transfer matrix $E$ in sector $q$.
This is due to the fact that
	the spectrum of the transfer matrix $E$ is split into different sectors.
	Actually, the correlation length $\xi$ may be identified as one of the $\xi_{q}$'s. Here we remark that, for a chosen value of the bond dimension $\chi$, one is able to produce
 the correlation length $\xi_{u,q}(\chi)$ and the correlation length $\xi_{q}(\chi)$ from the iMPS representation.

Physically,  the correlation length $\xi_{u,q}(\chi)$
	in sector $q$, is {\it temporal}, because it essentially describes a time evolution, due to the fact that {\it only}  one sector is involved.
This amounts to stating that $q$ is conserved. This is in contrast to the correlation length $\xi_{q}(\chi)$ which is {\it spatial}, since it involves
	different sectors such that a local spatial disturbance propagates across a given block,
	as far as the spread of information is concerned.
Remarkably, for a scale-invariant state, one may expect that both the correlation length $\xi_{u,q}(\chi)$ and the correlation length $\xi_{q}(\chi)$ must scale with the bond dimension $\chi$ in an identical way, regardless of the values of $q$. More precisely, one may anticipate that the scaling relations of  both the correlation length $\xi_{u,q}(\chi)$ and the correlation length $\xi_{q}(\chi)$ with the bond dimension $\chi$ in the iMPS representation do not depend on $q$.
As a consequence, it is sufficient to consider $\xi_{u}(\chi)$ and $\xi(\chi)$, due to their practical $q$ independence.

	Hence it is necessary to take into account the dynamical critical exponent $z$, as it appears in the dispersion relation $\omega\sim
	k^z$.
	As such, we might speculate that there is a remarkable scaling relation between $\xi(\chi)$ and $\xi_u(\chi)$, of the form
	\begin{equation}
		{\xi_u(\chi) \sim \xi^{1/z}(\chi)} \,.
		\label{C}
	\end{equation}
This offers an alternative means to extract the dynamical critical exponent $z$, which even holds for conformal field theory with $z=1$,
as a result of the Lorentz invariance.
That is, there is no difference between the two choices, i.e., either  $\xi(\chi)$ or $\xi_u(\chi)$, for conformally invariant states, if a scaling relation is concerned. In practice, this observation offers an alternative way to determine the correlation length $\xi$ from simulation results of the iDMRG algorithm when a proper symmetry group is implemented. In contrast, this is not the case for scale-invariant states arising from SSB with type-B GMs, given they are not subject to the Lorentz invariance.
Here, we remark that  the correlation length $\xi_u(\chi)$  has been exploited in Ref.~\cite{chenpc} to study the scaling behavior of the entanglement entropy $S_f(\chi)$. Instead,
the correlation length $\xi$ has been adopted in Ref.~\cite{dai}. This explains the discrepancy between the scaling relations there.

    As an illustration, we evaluate $\xi_{u,q}(\chi)$  for the spin-$s$ Heisenberg ferromagnetic model~(3), the $\rm{SU}(3)$ ferromagnetic model~(4) and the SO(4) spin-orbital model~(5). For our purpose, the iDMRG algorithm is the method of choice to simulate the models under investigation, since it is necessary to implement a proper symmetry group, i.e., the residual symmetry group $H$, to target a specific ground state~\cite{website}. Specifically, the rank $R$ is 1 for the spin-$s$ Heisenberg ferromagnetic model, and the rank $R$ is 2 for the $\rm{SU}(3)$ ferromagnetic model and the SO(4) spin-orbital model.
    Accordingly, the  residual
	symmetry group $\rm {U} (1)$ has been implemented for the spin-$s$ Heisenberg ferromagnetic model, and the  residual
	symmetry group $\rm {U} (1) \times \rm {U} (1)$ has been implemented for the $\rm{SU}(3)$ ferromagnetic model
    and the SO(4) spin-orbital model. In addition, we evaluate $\xi_{q}(\chi)$ for each of the three models.
	The dynamical critical exponent $z$ can be extracted
	according to the scaling relation (\ref{C}) between $\xi_u(\chi)$ and $\xi(\chi)$.

	\begin{figure}
		\includegraphics[width=0.47\textwidth]{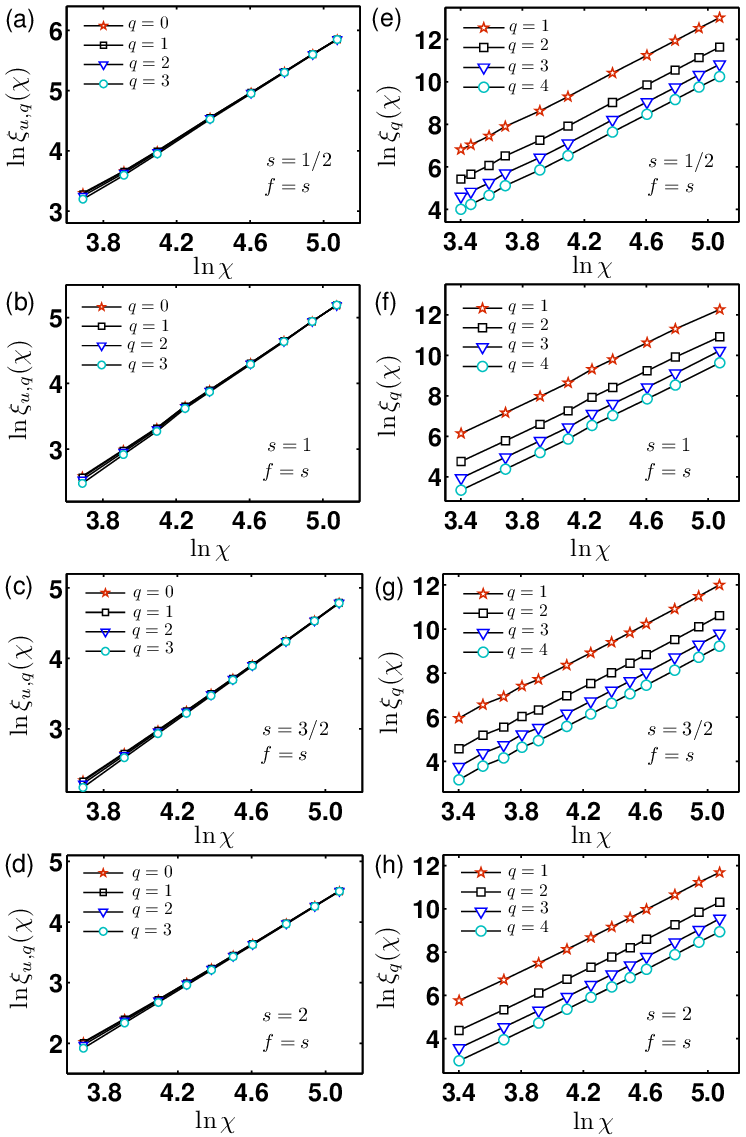}
		\caption{ The spin-$s$ Heisenberg ferromagnetic model with filling $f=s$. (Left panels) Scaling relation between the correlation
          length $\xi_{u,q}(\chi)$  and
			bond dimension $\chi$, with $\chi$ ranging from $40$ to $160$ and $q=0, 1, 2, 3$:
            (a)\;$s=1/2$; (b)\;$s=1$; (c)\; $s=3/2$; (d)\;$s=2$.
             (Right panels) Scaling relation between the correlation length $\xi_{q}(\chi)$ in sector $q$ and
			 bond dimension $\chi$, with $\chi$ ranging from $30$ to $160$ and $q= 1, 2, 3, 4$:
             (e)\;$s=1/2$; (f)\;$s=1$; (g)\;$s=3/2$; (h)\;$s=2$.}
		\label{q0Su2CorrU}
	\end{figure}


	\begin{figure}
		\includegraphics[width=0.47\textwidth]{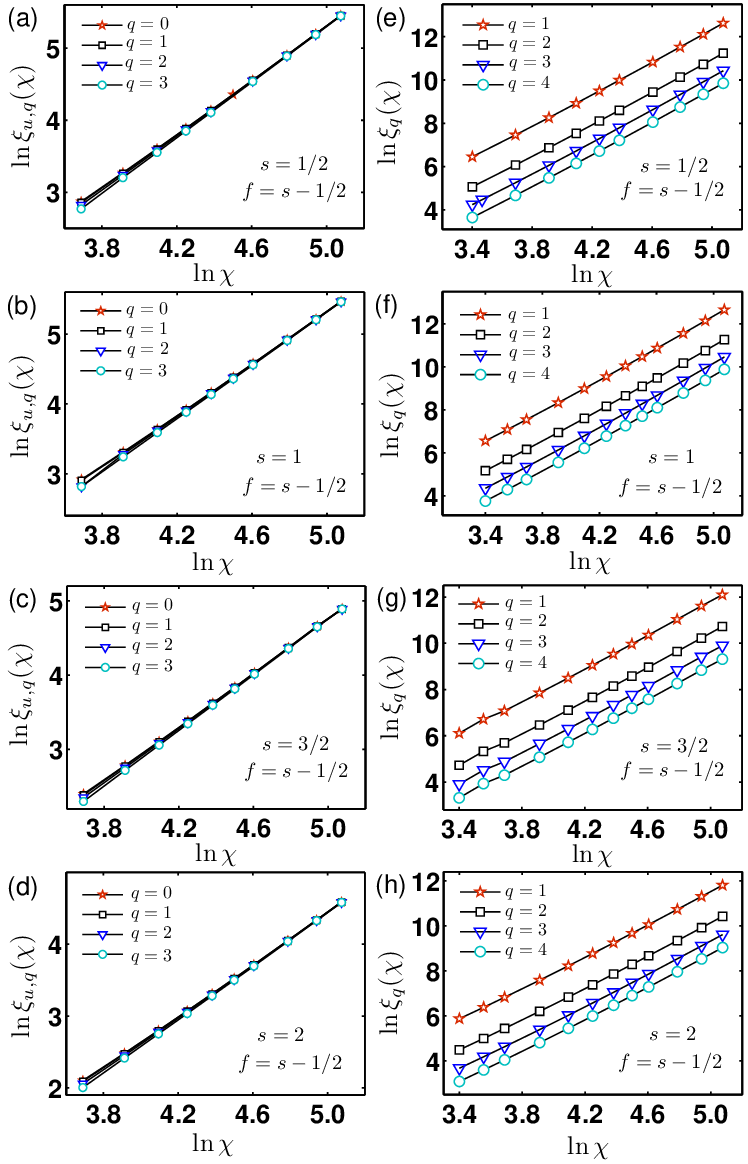}
		\caption{ The spin-$s$ Heisenberg ferromagnetic model with filling $f=s-1/2$. (Left panels) Scaling relation between the correlation length $\xi_{u,q}(\chi)$  and
			bond dimension $\chi$, with $\chi$ ranging from $40$ to $160$ and $q=0, 1, 2, 3$:
            (a)\;$s=1/2$; (b)\;$s=1$; (c)\;$s=3/2$; (d)\;$s=2$.
             (Right panels) Scaling relation between the correlation length $\xi_{q}(\chi)$  and
			bond dimension $\chi$, with $\chi$ ranging from $30$ to $160$ and $q= 1, 2, 3, 4$:
           (e)\;$s=1/2$; (f)\;$s=1$; (g)\;$s=3/2$; (h)\;$s=2$.}
 		\label{q1Su2CorrU}
	\end{figure}

	\subsubsection*{1.~ The spin-$s$ Heisenberg ferromagnetic model}
	
	For filling $f=s$, we plot the scaling of the correlation length $\xi_{u,q}(\chi)$ with bond dimension in Fig.~\ref{q0Su2CorrU}(a)(b)(c)(d) and
the scaling of the correlation length $\xi_q(\chi)$ with bond dimension in Fig.~\ref{q0Su2CorrU}(e)(f)(g)(h), for the spin-$s$ Heisenberg ferromagnetic model~(3), with  $s=1/2, 1, 3/2$ and $2$.
The corresponding plots for filling $f=s-1/2$ are shown in Fig.~\ref{q1Su2CorrU}.
Our numerical results confirm that the scaling relations for both correlation lengths $\xi_{u,q}(\chi)$ and $\xi_{q}(\chi)$, as follow from the iDMRG simulations in the iMPS representation, do not depend on $q$, within a reasonable error (less than $3.9\%$).
As a consequence, one only needs to introduce the two correlation lengths $\xi_u(\chi)$ and $\xi(\chi)$, as far as their scaling relation is concerned.

\begin{figure}
		\includegraphics[width=0.48\textwidth]{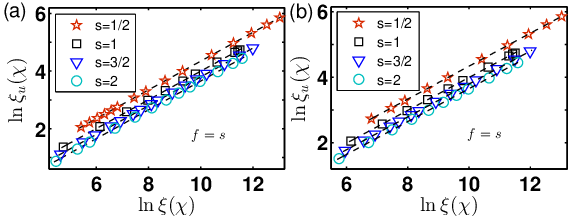}
		\caption{
	        Scaling relation between the correlation lengths $\xi_u(\chi)$ and $\xi(\chi)$ for the spin-$s$ Heisenberg ferromagnetic model
			with $s=1/2, 1, 3/2$ and $2$ and filling $f=s$. Different sectors are labelled by $q$.
			(a) The sector $q=0$, with bond dimension $\chi$ ranging from $20$ to $160$.
			(b) The sector $q=1$, with bond dimension $\chi$ ranging from $30$ to $160$.
			Corresponding estimates for the scaling exponent are shown in Table~\ref{dynamicz0}.}
		\label{FZq0}
	\end{figure}

\begin{figure}
		\includegraphics[width=0.48\textwidth]{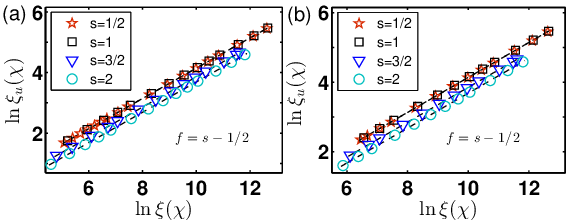}
		\caption{Scaling relation between the correlation lengths $\xi_u(\chi)$  and $\xi(\chi)$  for the spin-$s$ Heisenberg ferromagnetic
            model
			with $s=1/2, 1, 3/2$ and $2$ and filling $f=s-1/2$.
		(a) The sector $q=0$, with bond dimension $\chi$ ranging from $20$ to $160$.
		(b) The sector $q=1$, with bond dimension $\chi$ ranging from $30$ to $160$. 	
		Corresponding estimates for the scaling exponent are shown in Table~\ref{dynamicz1}. }
		\label{FZq1}
	\end{figure}

For this same model, in Fig.~\ref{FZq0} and  Fig.~\ref{FZq1}, we plot the scaling relation between the correlation lengths $\xi_u(\chi)$ and
$\xi(\chi)$ for $s=1/2, 1, 3/2$ and $2$, with different fillings $f=s$ and $f=s-1/2$.
 The scaling relation (\ref{C}) is confirmed, regardless of our choice of sector $q$, to evaluate the correlation length $\xi_u(\chi)$,  thus offering an alternative means to extract the dynamical critical exponent $z$.
The best linear fits yield $z=2$,
with relative error less than $1\%$, compared to the exact value $z=2$ from the conventional spin wave theory, as shown in
Table~\ref{dynamicz0} and Table~\ref{dynamicz1}.

	\begin{table}
		\renewcommand\arraystretch{2}
		\caption{Estimates for the dynamical critical exponent $z$, compared to the exact value $z=2$, for the spin-$s$ Heisenberg
         ferromagnetic model, with
			filling $f=s$.
			Results for sectors $q=0$ and $q=1$ are shown.
			}
		        \vskip 2mm
	       \begin{tabular}{cccccccc}
			\hline\hline
             \begin{minipage}{1cm} Sector \end{minipage}
			 &\begin{minipage}{0.6cm} $s$ \end{minipage} &
			\begin{minipage}{1.1cm} $1/2$ \end{minipage} &
			\begin{minipage}{1.1cm} $1$ \end{minipage} &
			\begin{minipage}{1.1cm} $3/2$ \end{minipage} &
			\begin{minipage}{1.1cm} $2$ \end{minipage}&\\
			\hline
			\begin{minipage}{1.2cm}  $q=0$ \end{minipage}
			&  & 1.9988 & 1.9948& 1.9992 & 2.0024  \\
			\begin{minipage}{1.2cm}  $q=1$ \end{minipage}
			&  & 2.0104 & 2.0052 & 2.0072 & 2.0076\\
			\hline\hline
		\end{tabular}
		\label{dynamicz0}
	\end{table}

	\begin{table}
		\renewcommand\arraystretch{2}
		\caption{Estimates for the dynamical critical exponent $z$, compared to the exact value $z=2$, for the spin-$s$ Heisenberg
         ferromagnetic model, with
			filling $f=s-1/2$.
			Results for sectors $q=0$ and $q=1$ are shown.
			}	
			\vskip 2mm
		\begin{tabular}{cccccccc}
			\hline\hline
         \begin{minipage}{1cm} Sector \end{minipage}
			& \begin{minipage}{0.6cm} $s$ \end{minipage} &
			\begin{minipage}{1.1cm} $1/2$ \end{minipage} &
			\begin{minipage}{1.1cm} $1$ \end{minipage} &
			\begin{minipage}{1.1cm} $3/2$ \end{minipage} &
			\begin{minipage}{1.1cm} $2$ \end{minipage}&\\
			\hline
			\begin{minipage}{1.2cm} $q=0$ \end{minipage}
			&   & 2.0004 & 1.9984 & 1.9972 & 2.0012  \\
			\begin{minipage}{1.2cm}  $q=1$ \end{minipage}
			&   & 2.0124 & 2.0096 & 2.006 & 2.0076\\
			\hline\hline
		\end{tabular}
		\label{dynamicz1}
	\end{table}

 \begin{figure}
		\includegraphics[width=0.47\textwidth]{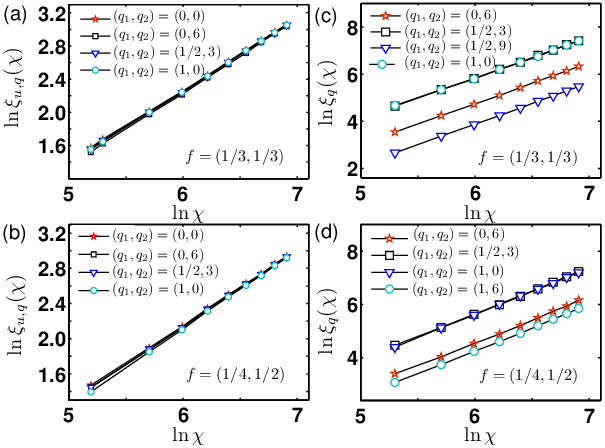}
		\caption{ The  $\rm{SU}(3)$ ferromagnetic model. (Left panels) Scaling relation between the correlation length $\xi_{u,q}(\chi)$ and
			bond dimension $\chi$, with $\chi$ ranging from $180$ to $1000$ and $(q_1, q_2)=(0, 0), (0, 6), (1/2, 3)$ and $(1, 0)$:
            (a)\;$f=(1/3, 1/3)$ and (b)\;$f=(1/4, 1/2)$.
             (Right panels) Scaling relation between the correlation length $\xi_{q}(\chi)$ and
			bond dimension $\chi$, with $\chi$ ranging from $200$ to $1000$:
             (c)\;$f=(1/3, 1/3)$ and (d)\;$f=(1/4, 1/4)$.
             The sector values $(q_1, q_2)$ are as indicated.}
		\label{SU3CorrU}
	\end{figure}

	\begin{figure}[ht]
		\includegraphics[width=0.48\textwidth]{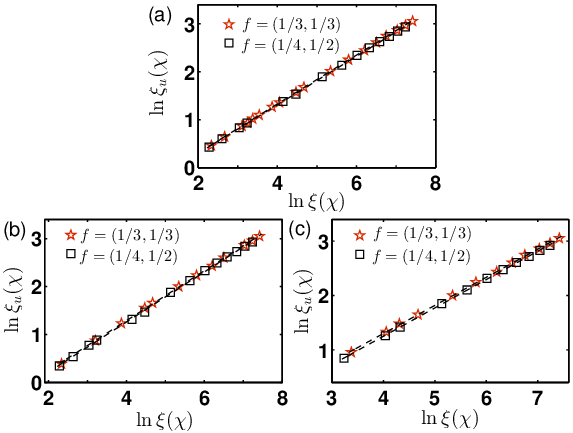}
		\caption{Scaling relation between the correlation lengths $\xi_u(\chi)$  and
			$\xi(\chi)$ for the $\rm{SU}(3)$ ferromagnetic model for different sectors,
			with fillings $f=(1/3,1/3)$ and $(1/4,1/2)$.
			(a) The sector $(q_1, q_2)=(0, 0)$, with bond dimension $\chi$ ranging from $50$ to $1000$.
			(b) The sector $(q_1, q_2)=(1/2, 3)$, with bond dimension $\chi$ ranging from $50$ to $1000$.
			(c) The sector $(q_1, q_2)=(1, 0)$, with bond dimension $\chi$ ranging from $90$ to $1000$.
			Corresponding estimates for the scaling exponent are shown in Table~\ref{dynamicz2}. }
		\label{FZSu3}
	\end{figure}

	\begin{table}[ht]
		\renewcommand\arraystretch{2}
		\caption{The dynamical critical exponent estimates for the $\rm{SU}(3)$ ferromagnetic model in
			sector $q$ with the indicated fillings.
			}
	        \vskip 2mm
		\begin{tabular}{cccccccc}
			\hline\hline
            \begin{minipage}{1cm} Sector \end{minipage}
			& \begin{minipage}{0.8cm} $f$ \end{minipage} &
			\begin{minipage}{1.5cm} $(1/3, 1/3)$ \end{minipage} &
			\begin{minipage}{1.5cm} $(1/4, 1/2)$ \end{minipage} &\\
			\hline
			\begin{minipage}{2.8cm} $(q_1, q_2)=(0, 0)$ \end{minipage}
			&  & 1.9589 &1.9739   \\
			\begin{minipage}{2.8cm}  $(q_1, q_2)=(1/2, 3)$ \end{minipage}
			&  & 1.9227& 1.9135\\
			\begin{minipage}{2.8cm}  $(q_1, q_2)=(1, 0)$ \end{minipage}
			&  & 1.9346& 1.9164\\
			\hline\hline
		\end{tabular}
		\label{dynamicz2}
	\end{table}

	\subsubsection*{2.~ The  $\rm{SU}(3)$ ferromagnetic model}
	
	Turning to the $\rm{SU}(3)$ ferromagnetic model~(4), for
	chosen fillings $f=(1/3, 1/3)$ and $f=(1/4, 1/2)$, we plot the scaling of the correlation length $\xi_{u,q}(\chi)$ with bond dimension in Fig.~\ref{SU3CorrU} (a)(b) and
	of correlation length $\xi_q(\chi)$ with bond dimension in Fig.~\ref{SU3CorrU}(c)(d).
	Our numerical results confirm that the scaling relations for both correlation lengths with the bond dimension, as follows from the iDMRG simulations in the iMPS representation, do not depend on $q$, within a reasonable error (less than $4\%$).
	As a consequence, one only needs to introduce the two correlation lengths $\xi_u(\chi)$ and $\xi(\chi)$, as far as their scaling relation is concerned.

   In Fig.~\ref{FZSu3}, we plot the scaling relation between the correlation lengths $\xi_u(\chi)$ and
  $\xi(\chi)$ with fillings $f=(1/3,1/3)$ and $(1/4,1/2)$.
  The scaling relation (\ref{C}) is again confirmed, regardless of our choice of sector to evaluate the correlation length $\xi_u(\chi)$, thus offering an alternative means to extract the dynamical critical exponent $z$.
	The best linear fits yield $z=2$, with a relative error less than $ 5\%$, compared to the exact value $z=2$ from the conventional spin wave theory, as shown in Table~\ref{dynamicz2}.

 \begin{figure}
		\includegraphics[width=0.47\textwidth]{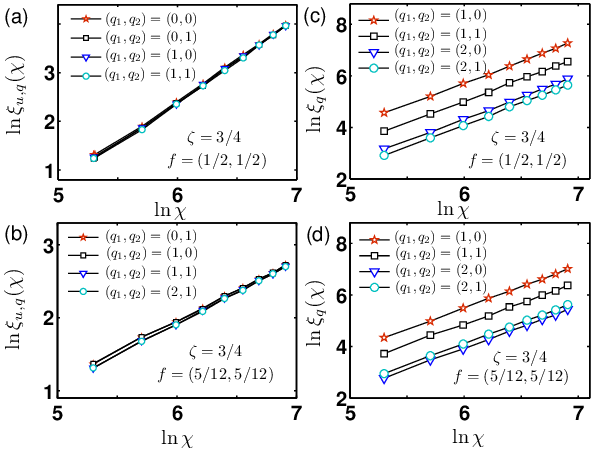}
		\caption{ The $\rm{SO}(4)$ spin-orbital model with $\zeta=3/4$.
		(Left panels) Scaling relation between the correlation length $\xi_{u,q}(\chi)$  and
			bond dimension $\chi$, with $\chi$ ranging from $200$ to $1000$ and indicated values for $(q_1,q_2)$:
            (a)\;$f=(1/2, 1/2)$ and (b)\;$f=(5/12, 5/12)$.
             (Right panels) Scaling relation between the correlation length $\xi_{q}(\chi)$  and
			bond dimension $\chi$, with $\chi$ ranging from $200$ to $1000$:
             (c)\;$f=(1/2, 1/2)$  and (d)\;$f=(5/12, 5/12)$.}
		\label{So4CorrU1}
	\end{figure}

\begin{figure}
		\includegraphics[width=0.47\textwidth]{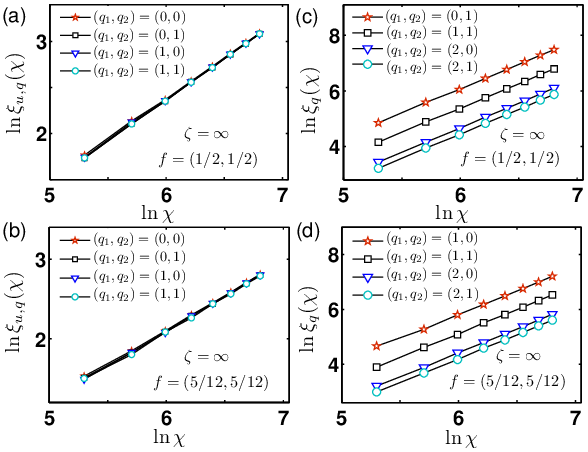}
		\caption{ The $\rm{SO}(4)$ spin-orbital model with $\zeta=\infty$.
		(Left panels) Scaling relation between the correlation length $\xi_{u,q}(\chi)$ and
			bond dimension $\chi$, with $\chi$ ranging from $200$ to $900$ and indicated values for $(q_1,q_2)$:
			(a)\;$f=(1/2, 1/2)$ and (b)\;$f=(5/12, 5/12)$.
             (Right panels) Scaling relation between the correlation length $\xi_{q}(\chi)$  and
			bond dimension $\chi$, with $\chi$ ranging from $200$ to $900$:
             (c)\;$f=(1/2, 1/2)$ and (d)\;$f=(5/12, 5/12)$.}
		\label{So4CorrU2}
	\end{figure}


	\begin{figure}[ht]
		\includegraphics[width=0.48\textwidth]{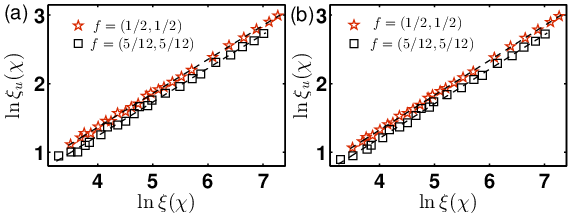}
		\caption{Scaling relation between the correlation lengths $\xi_u(\chi)$ and
			$\xi(\chi)$ for the $\rm{SO}(4)$ spin-orbital model with $\zeta=3/4$.
  			(a) The sector $(q_1, q_2)=(0, 0)$ with bond dimension $\chi$ ranging from $100$ to $1000$.
            (b) The sector $(q_1, q_2)=(1, 0)$ with bond dimension $\chi$ ranging from $100$ to $1000$.
			The fillings are chosen to be $f=(1/2,1/2)$ and $(5/12,5/12)$.
			The corresponding estimates for the dynamical critical exponent $z$ are given in Table~\ref{dynamicz3}.}
		\label{FZAklt}
	\end{figure}
	
	\begin{figure}
		\includegraphics[width=0.48\textwidth]{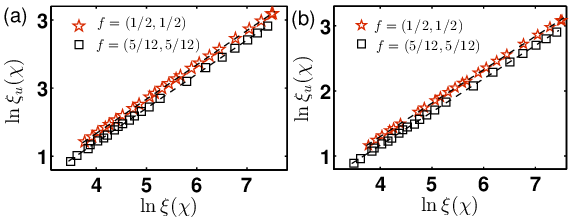}
		\caption{Scaling relation between the correlation lengths $\xi_u(\chi)$ and
			$\xi(\chi)$ for the $\rm{SO}(4)$ spin-orbital model with $\zeta=\infty$.
			(a) The sector $(q_1, q_2)=(0, 0)$ with bond dimension $\chi$ ranging from $100$ to $1000$.
            (b) The sector $(q_1, q_2)=(1, 0)$ with bond dimension $\chi$ ranging from $100$ to $1000$.
			The fillings are chosen to be $f=(1/2,1/2)$ and $(5/12,5/12)$.
			The corresponding estimates for the dynamical critical exponent $z$ are given in Table~\ref{dynamicz4}.}
		\label{FZFerro}
	\end{figure}

	\begin{table}[ht]
		\renewcommand\arraystretch{2}
		\caption{Estimates of the dynamical critical exponent $z$ for the $\rm{SO}(4)$ spin-orbital model with $\zeta=3/4$ for sectors
       $(q_1, q_2)=(0, 0)$ and $(q_1, q_2)=(1, 0)$.
		The bond dimension ranges from $100$ to $1000$, with a relative error less than $2\%$, compared to the exact value $z=2$.
	        The fillings $f$ are as indicated.}
	        \vskip 2mm
		\begin{tabular}{cccccccc}
			\hline\hline
            \begin{minipage}{1cm} Sector \end{minipage}
			& \begin{minipage}{0.8cm} $f$ \end{minipage} &
			\begin{minipage}{1.5cm} $(1/2, 1/2)$ \end{minipage} &
			\begin{minipage}{1.8cm} $(5/12, 5/12)$ \end{minipage} &\\
			\hline
			\begin{minipage}{2.5cm} $(q_1, q_2)=(0, 0)$ \end{minipage}
			&  & 2.0214 &  2.0288 \\
			\begin{minipage}{2.5cm}  $(q_1, q_2)=(1, 0)$ \end{minipage}
			&  & 1.9728 & 1.99\\
			\hline\hline
		\end{tabular}
		\label{dynamicz3}
	\end{table}

	\begin{table}[ht]
		\renewcommand\arraystretch{2}
		\caption{Estimates of the dynamical critical exponent $z$ for the $\rm{SO}(4)$ spin-orbital model with $\zeta=\infty$ for sectors $(q_1, q_2)=(0, 0)$ and $(q_1, q_2)=(1, 0)$.
		The bond dimension ranges from $100$ to $1000$, with a relative error less than $3\%$, compared to the exact value $z=2$.
	        The fillings $f$ are as indicated.}
	        \vskip 2mm
		\begin{tabular}{ccccccccccccc}
			\hline\hline
          \begin{minipage}{1cm} Sector \end{minipage}
			& \begin{minipage}{0.8cm} $f$ \end{minipage} &
			\begin{minipage}{1.5cm} $(1/2, 1/2)$ \end{minipage} &
			\begin{minipage}{1.8cm} $(5/12, 5/12)$ \end{minipage} &\\
			\hline
			\begin{minipage}{2.5cm} $(q_1, q_2)=(0, 0)$ \end{minipage}
			&  & 1.9853 &  1.9976 \\
			\begin{minipage}{2.5cm}  $(q_1, q_2)=(1, 0)$ \end{minipage}
			&  & 1.9497 & 1.9585\\
			\hline\hline
       		\end{tabular}
		\label{dynamicz4}
	\end{table}

	\subsubsection*{3.~ The $\rm{SO}(4)$ spin-orbital model}
	
For the $\rm{SO}(4)$ spin-orbital model~(5) with $\zeta=3/4$ and $\zeta=\infty$, and chosen fillings $f=(1/2, 1/2)$ and $(5/12, 5/12)$, we plot the scaling of the correlation length $\xi_{u,q}(\chi)$ with bond dimension in Fig.~\ref{So4CorrU1}(a)(b) and ~\ref{So4CorrU2}(a)(b) and
	the correlation length $\xi_q(\chi)$ with the bond dimension $\chi$ in Fig.~\ref{So4CorrU1}(c)(d) and ~\ref{So4CorrU2}(c)(d).
	Our numerical results for this model also confirm that the scaling relations for both correlation lengths $\xi_{u,q}(\chi)$ and $\xi_{q}(\chi)$ with bond dimension, as follow from the iDMRG simulations in the iMPS representation, do not depend on $q$, within a reasonable error (less than $3.5\%$).
	We thus only consider the two correlation lengths $\xi_u(\chi)$ and $\xi(\chi)$, as far as their scaling relation is concerned.

	In Fig.~\ref{FZAklt} and Fig.~\ref{FZFerro}, we plot the scaling relation between the correlation lengths $\xi_u(\chi)$ and
    $\xi(\chi)$ with $\zeta=3/4$ and $\zeta=\infty$, respectively.
    The scaling relation (\ref{C}) is again confirmed, regardless of our choice of a sector to evaluate the correlation length $\xi_u(\chi)$,  thus offering an alternative means to extract the dynamical critical exponent $z$. In this case, the best linear fits yield $z=2$,
	with the relative error less than $2\%$ and $3\%$, compared to the exact value $z=2$ from the conventional spin wave theory.
	The estimates are shown in Table~\ref{dynamicz3} and Table~\ref{dynamicz4}.

	\section*{D.~~ The number of type-B GMs extracted from finite block-size scaling with increasing $\chi$}
	\label{appB}
	
	As follows from a finite block-size scaling analysis of the entanglement entropy $S_f(n)$~\cite{FMGM0,golden0} given in Eq.~(1), the number of type-B GMs may be extracted from the iDMRG simulations~\cite{idmrg0} for different values of the bond dimension $\chi$, if a proper symmetry is implemented.
	In principle, one may perform an extrapolation to obtain an accurate estimate of the number of type-B GMs, as the bond dimension $\chi$ tends to infinity.
	\begin{figure}
		\includegraphics[width=0.44\textwidth]{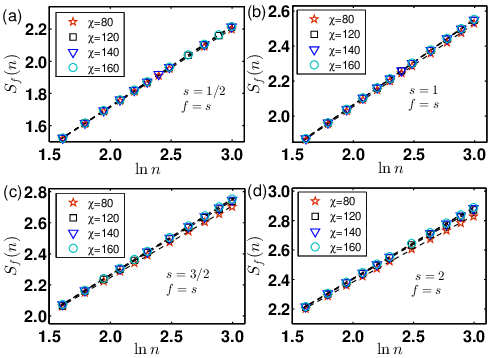}
		\caption{The entanglement entropy $S_f(n)$ versus $\ln n$ for the $\rm {SU} (2)$ spin-$s$ Heisenberg ferromagnetic model~(3) with filling $f=s$, where
			(a) $s=1/2$, (b) $s=1$, (c) $s=3/2$ and (d) $s=2$. The bond dimensions are $\chi=80, 120, 140$ and $160$, with the
			block size ranging from $n=5$ to $n=20$.}
		\label{FIGSandNq0}
	\end{figure}
	
	\begin{table}
		\renewcommand\arraystretch{2}
		\caption{Estimates for the number of type-B GMs $N_B$ extracted from the finite block-size scaling of the entanglement entropy
			$S_f(n)$
			for the $\rm {SU} (2)$ spin-$s$ Heisenberg ferromagnetic model~(3) with filling $f=s$ and bond dimensions $\chi=80, 120,
           140, 160$.
			The block size ranges from $n=5$ to $n=20$.}
		\vskip 2mm
		\begin{tabular}{cccccccc}
			\hline
			\hline
			& \begin{minipage}{1cm} $\chi$ \end{minipage} &
			\begin{minipage}{1.1cm} $80$ \end{minipage} &
			\begin{minipage}{1.1cm} $120$ \end{minipage} &
			\begin{minipage}{1.1cm} $140$ \end{minipage} &
			\begin{minipage}{1.1cm} $160$ \end{minipage}&\\
			\hline
			\begin{minipage}{1.2cm}$s=1/2$ \end{minipage}
			& $N_B$ & 0.9804 & 0.9946 & 0.9986 & 1.0002 \\
			\begin{minipage}{1.2cm}$s=1$ \end{minipage}
			&  $N_B$ & 0.9534 & 0.9758 & 0.978 & 0.982\\
			\begin{minipage}{1.2cm}$s=3/2$ \end{minipage}
			&  $N_B$ &0.9248 & 0.9668 & 0.9722 & 0.9786\\
			\begin{minipage}{1.2cm} $s=2$ \end{minipage}
			&$N_B$ & 0.9054 & 0.9156 & 0.9632 & 0.97\\
			\hline
			\hline
		\end{tabular}
		\label{SandNq0}
	\end{table}

	\begin{figure}
		\includegraphics[width=0.44\textwidth]{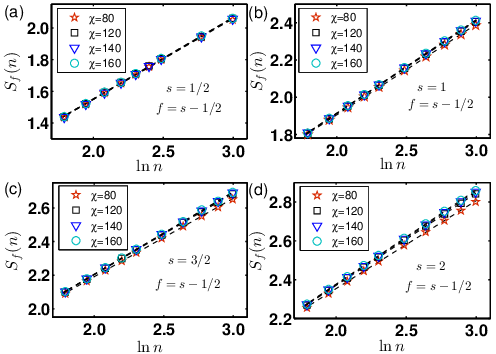}
		\caption{The entanglement entropy $S_f(n)$ versus $\ln n$ for the $\rm {SU} (2)$ spin-$s$ Heisenberg ferromagnetic model~(3) with filling $f=s-1/2$, where
			(a) $s=1/2$, (b) $s=1$, (c) $s=3/2$ and (d) $s=2$. The bond dimensions are $\chi=80, 120, 140, 160$.
			The block size ranges from $n=6$ to $n=20$.}
		\label{FIGSandNq1}
	\end{figure}

	\begin{table}
		\renewcommand\arraystretch{2}
		\caption{Estimates for the number of type-B GMs $N_B$ extracted from a finite block-size scaling of the entanglement entropy
			$S_f(n)$ for the $\rm {SU} (2)$ spin-$s$ Heisenberg ferromagnetic model~(3) with filling $f=s-1/2$ and bond dimensions $\chi=80,
           120, 140,  160$.
			The block size ranges from $n=6$ to $n=20$.}
		\vskip 2mm
		\begin{tabular}{cccccccc}
			\hline
			\hline
			& \begin{minipage}{1cm} $\chi$ \end{minipage} &
			\begin{minipage}{1.1cm} $80$ \end{minipage} &
			\begin{minipage}{1.1cm} $120$ \end{minipage} &
			\begin{minipage}{1.1cm} $140$ \end{minipage} &
			\begin{minipage}{1.1cm} $160$ \end{minipage}&\\
			\hline
			\begin{minipage}{1.2cm}$s=1/2$ \end{minipage}
			& $N_B$ & 1.0214 & 1.0278 & 1.0288 & 1.0298 \\
			\begin{minipage}{1.2cm}$s=1$ \end{minipage}
			&  $N_B$ & 0.9732 & 0.9962 & 0.999 & 1.0052\\
			\begin{minipage}{1.2cm}$s=3/2$ \end{minipage}
			&  $N_B$ &0.931 & 0.9676 & 0.9756 & 0.9804\\
			\begin{minipage}{1.2cm} $s=2$ \end{minipage}
			&$N_B$ & 0.9064 & 0.9526 & 0.9642 & 0.9722\\
			\hline
			\hline
		\end{tabular}
		\label{SandNq1}
	\end{table}

	\subsubsection*{1.~ The $\rm{SU}(2)$ spin-$s$ Heisenberg ferromagnetic model}
	
	Fig.~\ref{FIGSandNq0} shows a plot of the entanglement entropy $S_f(n)$ versus $\ln n$ for the $\rm {SU} (2)$ spin-$s$ Heisenberg ferromagnetic model~(3) with filling $f=s$ and different values of $s$. The best linear fit yields $N_B=1$, with relative error less than $3\%$, with data shown in Table \ref{SandNq0}.

	Fig.~\ref{FIGSandNq1} shows a plot of the entanglement entropy $S_f(n)$ versus $\ln n$ for the $\rm {SU} (2)$ spin-$s$ Heisenberg ferromagnetic model~(3) with filling $f=s-1/2$ and different values of $s$.
	The best linear fit yields $N_B=1$, with a relative error less than $3\%$, as indicated in Table \ref{SandNq1}.

\begin{figure}
		\includegraphics[width=0.46\textwidth]{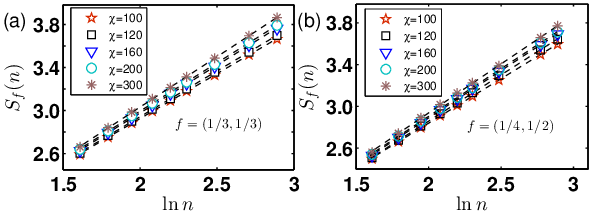}
		\caption{The  entanglement entropy $S_f(n)$ versus $\ln n$ for the $\rm{SU}(3)$ ferromagnetic model~(4)
			for filling values
			(a) $f=(1/3, 1/3)$ and (b) $f=(1/4, 1/2)$. The bond dimensions are $\chi=100, 120, 160, 200, 300$, with the block
			size ranging from $n=5$ to $n=18$.}
		\label{FIGSandNSU3}
	\end{figure}

	\begin{table}
		\renewcommand\arraystretch{2}
		\caption{Estimates for the number of type-B GMs $N_B$ extracted from the finite block-size scaling of the entanglement entropy
			$S_f(n)$ for the $\rm{SU}(3)$ ferromagnetic model~(4) with fillings $f=(1/3, 1/3)$ and $f=(1/4, 1/2)$. The bond
            dimensions are
			$\chi=100, 120, 160, 200, 300$, with the block size $n$ ranging from $n=5$ to $n=18$.
		}
		\vskip 2mm
		\begin{tabular}{cccccccc}
			\hline
			\hline
			&\begin{minipage}{0.6cm} $\chi$ \end{minipage} &
			\begin{minipage}{0.9cm} $100$ \end{minipage} &
			\begin{minipage}{0.9cm} $120$ \end{minipage} &
			\begin{minipage}{0.9cm} $160$ \end{minipage} &
			\begin{minipage}{0.9cm} $200$ \end{minipage} &
			\begin{minipage}{0.9cm} $300$ \end{minipage}&\\
			\hline
			\begin{minipage}{2cm} $f=(1/3, 1/3)$ \end{minipage}
			&$N_B$  & 1.6686 & 1.6982 & 1.7732 & 1.791 & 1.8702 &\\
			\begin{minipage}{2cm}$f=(1/4, 1/2)$ \end{minipage}
			&$N_B $& 1.704 & 1.771 & 1.8138 &1.825  & 1.8772 &\\
			\hline
			\hline
		\end{tabular}
		\label{SandNqsu3}
	\end{table}

	
	\subsubsection*{2.~  The  $\rm{SU}(3)$ ferromagnetic model}
	
	Fig.~\ref{FIGSandNSU3} shows a plot of the entanglement entropy $S_f(n)$ versus $\ln n$ for the $\rm{SU}(3)$ ferromagnetic model~(4) for different fillings.
		The best linear fit yields $N_B=2$,  with relative error less than $6.5\%$, as indicated in Table \ref{SandNqsu3}.
	
		\begin{figure}[h]
		\includegraphics[width=0.45\textwidth]{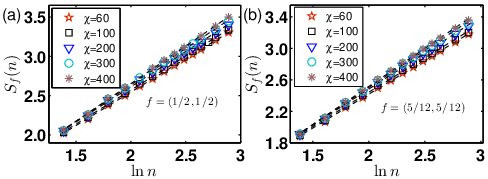}
		\caption{The entanglement entropy $S_f(n)$ versus $\ln n$ for the $\rm{SO}(4)$ spin-orbital model~(5) with  $\zeta=3/4$ for
			 fillings (a) $f=(1/2, 1/2)$ and (b) $f=(5/12, 5/12)$. The bond dimensions are $\chi=60, 100, 200, 300, 400$, with block size
			ranging from $n=4$ to $n=18$.		}
		\label{SandNso4AKLT}
	\end{figure}

	\begin{table}
		\renewcommand\arraystretch{2}
		\caption{Estimates for the number of type-B GMs $N_B$ extracted from the finite block-size scaling of the entanglement entropy
			$S_f(n)$ for the $\rm{SO}(4)$ spin-orbital model~(5) with  $\zeta=3/4$.
			The fillings  are $f=(1/2, 1/2)$ and $f=(5/12, 5/12)$, where the bond dimensions are $\chi=60, 100, 200, 300, 400$, with block
            size ranging from $n=4$ to $n=18$.}
				\vskip 2mm
		\begin{tabular}{cccccccc}
			\hline
			\hline
			&\begin{minipage}{0.6cm} $\chi$ \end{minipage} &
			\begin{minipage}{0.9cm} $60$ \end{minipage} &
			\begin{minipage}{0.9cm} $100$ \end{minipage} &
			\begin{minipage}{0.9cm} $200$ \end{minipage} &
			\begin{minipage}{0.9cm} $300$ \end{minipage} &
			\begin{minipage}{0.9cm} $400$ \end{minipage}&\\
			\hline
			\begin{minipage}{2.4cm} $f=(1/2, 1/2)$ \end{minipage}
			&$N_B$  & 1.703 & 1.7302 & 1.7886 & 1.8516 & 1.9104 &\\
			\begin{minipage}{2.4cm}$f=(5/12, 5/12)$ \end{minipage}
			&$N_B $& 1.7236 & 1.7466 & 1.8358 & 1.8906 & 1.9352 &\\
			\hline
			\hline
		\end{tabular}
		\label{SandNqso4a}
	\end{table}

\subsubsection*{3.~  The  $\rm{SO}(4)$ spin-orbital model}
	
Fig.~\ref{SandNso4AKLT} shows a plot of the entanglement entropy $S_f(n)$ versus $\ln n$ for the $\rm{SO}(4)$ spin-orbital model~(5) with  $\zeta=3/4$, for different fillings. The best linear fit yields $N_B=2$, with relative error less than $4.5\%$, as indicated in Table~\ref{SandNqso4a}.
		
\begin{figure}
		\includegraphics[width=0.45\textwidth]{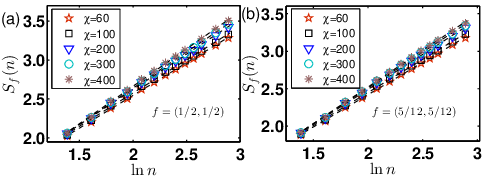}
		\caption{
			The  entanglement entropy $S_f(n)$ versus $\ln n$ for the $\rm{SO}(4)$ spin-orbital model~(5) with $\zeta=\infty$.
			The fillings are (a) $f=(1/2, 1/2)$ and (b) $f=(5/12, 5/12)$, with bond dimensions $\chi=60, 100, 200, 300, 400$, and block
			size ranging from $n=4$ to $n=18$.
		}
		\label{SandNso4Ferro}
	\end{figure}

Fig.~\ref{SandNso4Ferro} shows a plot of the entanglement entropy $S_f(n)$ versus $\ln n$ for the $\rm{SO}(4)$ spin-orbital model~(5) with  $\zeta=\infty$, for different fillings. The best linear fit yields $N_B=2$, with relative error less than $4\%$, as indicated in Table~\ref{SandNqso4b}.

\vspace{30pt}

In summary, our results for the three illustrative models indicate that the estimates for the number of type-B GMs $N_B$ tend to saturate, as the bond dimension $\chi$ increases. As a consequence, we are led to conclude that the number of type-B GMs $N_B$ may be reliably extracted from finite block-size scaling, as long as the bond dimension $\chi$ is large enough, within a reasonable accuracy.

	\begin{table}[htbp]
		\renewcommand\arraystretch{2}
		\caption{Estimates for the number of type-B GMs $N_B$ extracted from the finite block-size scaling of the entanglement entropy
			$S_f(n)$ for the $\rm{SO}(4)$ spin-orbital model~(5) with $\zeta=\infty$.
			The fillings are $f=(1/2, 1/2)$ and $f=(5/12, 5/12)$, with bond dimensions $\chi=60, 100, 200, 300, 400$, and block size
			ranging from $n=4$ to $n=18$.}
		\vskip 2mm		
		\begin{tabular}{cccccccc}
			\hline
			\hline
			&\begin{minipage}{0.6cm} $\chi$ \end{minipage} &
			\begin{minipage}{0.9cm} $60$ \end{minipage} &
			\begin{minipage}{0.9cm} $100$ \end{minipage} &
			\begin{minipage}{0.9cm} $200$ \end{minipage} &
			\begin{minipage}{0.9cm} $300$ \end{minipage} &
			\begin{minipage}{0.9cm} $400$ \end{minipage}&\\
			\hline
			\begin{minipage}{2.4cm} $f=(1/2, 1/2)$ \end{minipage}
			&$N_B$  & 1.6768 & 1.7058 & 1.813 & 1.8558 & 1.9212 &\\
			\begin{minipage}{2.4cm}$f=(5/12, 5/12)$ \end{minipage}
			&$N_B $& 1.7232 & 1.7682 &1.8662  & 1.9118 & 1.9544 &\\
			\hline
			\hline
		\end{tabular}
		\label{SandNqso4b}
	\end{table}
